# Superior mechanical flexibility of phosphorene and few-layer black phosphorus

Qun Wei,[1,2] Xihong Peng,[1*]

[1]School of Letters and Sciences, Arizona State University, Mesa, Arizona 85212, USA

[2]School of Physics and Optoelectronic Engineering, Xidian University, Xi'an, 710071, P.R. China

## ABSTRACT

Recently fabricated two dimensional (2D) phosphorene crystal structures have demonstrated great potential in applications of electronics. Mechanical strain was demonstrated to be able to significantly modify the electronic properties of phosphorene and few-layer black phosphorus. In this work, we employed first principles density functional theory calculations to explore the mechanical properties of phosphorene, including ideal tensile strength and critical strain. It was found that a monolayer phosphorene can sustain tensile strain up to 27% and 30% in the zigzag and armchair directions, respectively. This enormous strain limit of phosphorene results from its unique puckered crystal structure. We found that the tensile strain applied in the armchair direction stretches the pucker of phosphorene, rather than significantly extending the P-P bond lengths. The compromised dihedral angles dramatically reduce the required strain energy. Compared to other 2D materials such as graphene, phosphorene demonstrates superior flexibility with an order of magnitude smaller Young's modulus. This is especially useful in practical large-magnitude-strain engineering. Furthermore, the anisotropic nature of phosphorene was also explored. We derived a general model to calculate the Young's modulus along different directions for a 2D system.

**Keywords:** 2D phosphorene, few-layer black phosphorus, strain-stress relation, ideal strength, critical strain, Poisson's ratio, Young's modulus, shear modulus

Recently researchers have successfully fabricated a new 2D few-layer black phosphorus [1-4] and found that this material is chemically inert and has great transport properties. It was reported that it has a carrier mobility up to 1000 cm$^2$/V·s [1] and an on/off ratio up to 10$^4$ [2] was achieved for the phosphorene transistors at room temperature. In addition, this material shows a finite direct band gap at the Γ point of Brillouin zone [1, 2, 5-8] (in contrast to the vanishing gap in graphene), which opens doors for additional applications in optoelectronics.

Tailoring electronic properties of semiconductor nanostructures has been critical for applications in electronics. Strain has a long history of being used to tune electronic properties of semiconductors [9-17]. Adventitious strain is almost unavoidable experimentally, but more interesting cases come from intentionally introduced and controlled strains. One of the most prominent examples is the greatly enhanced mobility in the strained Si nanochannel [18, 19]. The approaches introducing strain include lattice mismatch, functional wrapping [20, 21], doping of material [22, 23], and direct mechanical application [14]. Researchers found that nanostructures maintain integrity under much higher strain than their bulk counterparts [16, 24], which dramatically expands the applicable strain in tuning properties of nanomaterials.

Intuitively, strain affects the electronic properties of materials only in a fine tuning manner. However, this is not always the case. As shown in recent studies [8, 15, 25], depending on the symmetry and bonding/antibonding nature of a specific electron orbital, the associated energy level can respond very differently to an applied strain. A qualitatively different band structure, such as, direct to indirect band gap transition [8, 15, 25] or directional preference of charge transport [26], may be obtained with a moderate strain.

In particular, 2D layered materials, such as graphene and MoS$_2$, possess great mechanical flexibility and can sustain a large strain (≥ 25%) [27-29]. In this work, we present detailed systematic analysis on the mechanical properties of monolayer phosphorene and few-layer black phosphorus. We found this material is super flexible and can withstand a tensile strain up to 30% and 32% for a monolayer and multi-layer of black phosphorus, respectively. Moreover, the Young's modulus of phosphorene in the armchair direction is significantly smaller than that of graphene, due to its unique puckered crystal structure. This makes the material very favorable in practical large-magnitude-strain engineering. Due to the extraordinary anisotropic nature of phosphorene, the Young's modulus was found sensitively dependent on the direction.

The theoretical calculations were carried out using first principles density functional theory (DFT) [30]. The Perdew-Burke-Ernzerhof (PBE) exchange-correlation functional [31] along with the projector-augmented wave (PAW) potentials [32, 33] were employed for the self-consistent total energy calculations and geometry optimization. The calculations were performed using the Vienna Ab-initio Simulation Package (VASP) [34, 35]. The kinetic energy cutoff for the plane wave basis set was chosen to be 350 eV. The reciprocal space was meshed at $14 \times 10 \times 1$ using Monkhorst-Pack method. The energy convergence criteria for electronic and ionic iterations were set to be $10^{-5}$ eV and $10^{-4}$ eV, respectively. With this parameter setting, the calculations were converged within 5 meV in total energy per atom. To simulate a monolayer and few-layer black phosphorus, a unit cell with periodic boundary condition was used. A vacuum space of at least 16 Å was included in the unit cell to minimize the interaction between the system and its replicas resulting from the periodic boundary condition.

The initial structures of monolayer and few-layer phosphorene were obtained from bulk black phosphorus [36]. Monolayer black phosphorus has a puckered honeycomb structure with each phosphorus atom covalently bonded with three adjacent atoms, as shown in Figure 1(a). Two-layer phosphorene has a sequence of AB stacking as bulk black phosphorus. The three- and four-layer phosphorene structures take the stacking sequence of ABA, and ABAB, respectively. Our calculated lattice constants for bulk black phosphorus are $a = 3.308$ Å, $b = 4.536$ Å, and $c = 11.099$ Å, in good agreement with experimental values [36] and other theoretical calculations [2, 37]. The relaxed lattice constants for a monolayer of phosphorene are $a = 3.298$ Å, $b = 4.627$ Å.

Starting with the relaxed phosphorene structures, tensile strain up to 45% was applied in either the x (zigzag) or y (armchair) direction to explore its ideal tensile strength (the highest achievable strength of a defect-free crystal at 0 K) [38, 39] and critical strain (the strain at which ideal strength reaches). Figure 1(c) and (d) present the schematics of the applied tensile strains. The tensile strain is defined as $\varepsilon = \dfrac{a - a_0}{a_0}$, where $a$ and $a_0$ are the lattice constants of the strained and relaxed structure, respectively. With each axial strain applied, the lattice constant in the transverse direction was fully relaxed through the technique of energy minimization to ensure the force in the transverse direction is a minimum.

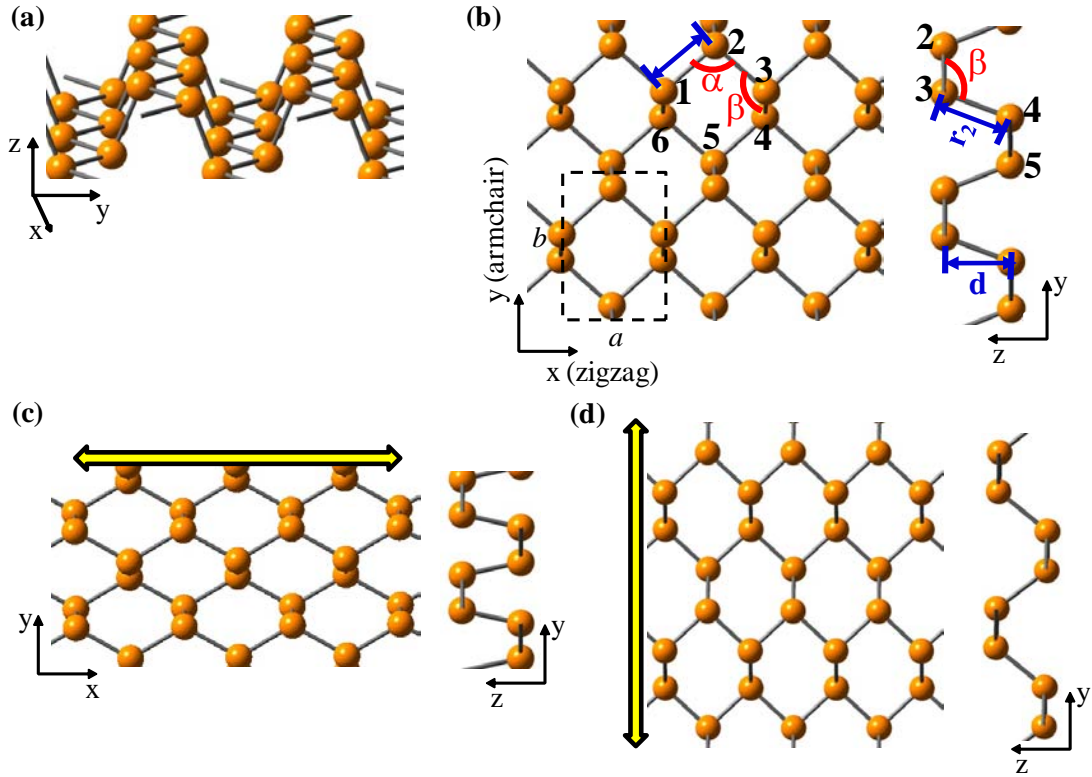

**Figure 1** Snapshots of a monolayer phosphorene structure. In (b), the dashed rectangle indicates a unit cell. Atoms were labeled using the numbers 1 through 6. Bond lengths $r_1$ and $r_2$, bond angles $\alpha$ and $\beta$ are denoted. The distance between the puckered layers in the z direction is labeled as d. (c) and (d) present the geometry of phosphorene under 30% axial tensile strain applied in the zigzag and armchair directions, respectively.

To calculate the ideal strength of 2D monolayer and few-layer phosphorene systems, we calculated their strain-stress relation using the method described in the references [40, 41]. This method was originally introduced for three dimensional crystals. To validate our calculations, we computed the mechanical properties of bulk black phosphorus, such as bulk modulus and elastic stiffness constants. Our calculated results are in good agreement with available experimental values. For example, our computed elastic constants are $C_{11}=189\,\text{GPa}$, $C_{22}=58\,\text{GPa}$, and $C_{33}=52\,\text{GPa}$. Experimental measured values are $C_{11}=179\,\text{GPa}$, $C_{22}=55\,\text{GPa}$, and $C_{33}=54\,\text{GPa}$.[42] Our predicted bulk modulus is 44 GPa. The experimental value is 33 GPa [43].

In a 2D system, the stress calculated from the DFT has to be modified to avoid the force being averaged over the entire simulation cell including the vacuum space. In order to compare directly with experiments and other calculations, we rescaled the stress by $Z/d_0$ to obtain the equivalent stress, where $Z$ is the cell length in the z direction and $d_0$ is the effective thickness of the system. For example, $d_0$ takes the interlayer spacing 5.55 Å of black phosphorus for the monolayer phosphorene. For the two-layer phosphorene, $d_0 = 11.10$ Å. And for the three- and four-layer phosphorene structures, $d_0$ takes the values of 16.65 Å and 22.20 Å, respectively.

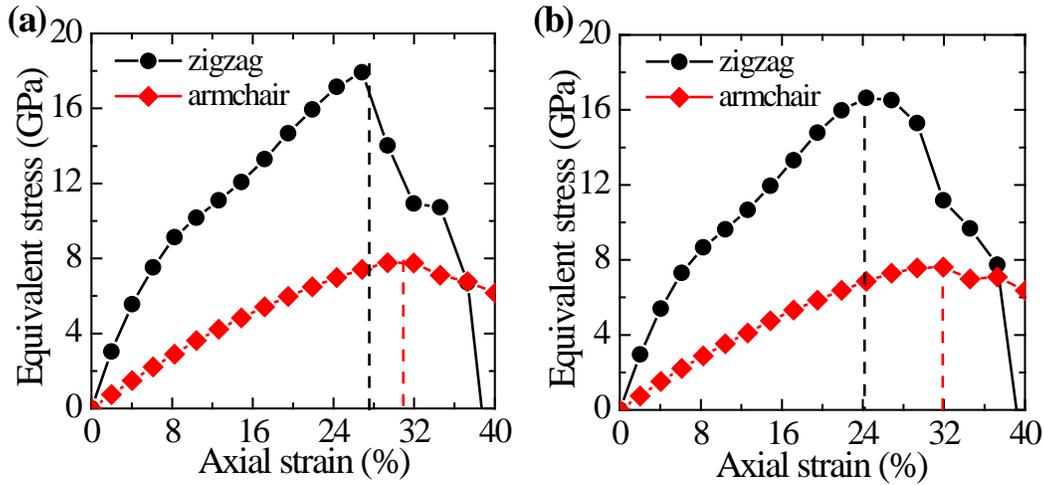

**Figure 2, The strain-stress relation for (a) monolayer and (b) two-layer phosphorene structures. The monolayer can sustain stress up to 18 GPa and 8 GPa in the zigzag and armchair directions, respectively. The corresponding critical strains are 27% (zigzag) and 30% (armchair). The ideal strengths for the multi-layer phosphorene are 16 GPa and 7.5 GPa in the zigzag and armchair directions, respectively, and their critical strains are 24% (zigzag) and 32% (armchair).**

Our calculated strain-stress relation is presented in Figure 2. For the monolayer phosphorene in Figure 2(a), the ideal strengths are 18 GPa and 8 GPa in the zigzag and armchair directions, respectively. The corresponding critical strains are 27% (zigzag) and 30% (armchair) [8]. In Figure 2(b), the two-layer phosphorene can withstand a stress up to 16 GPa and 7.5 GPa in the x and y directions, respectively, The corresponding critical strains are 24% (zigzag) and 32% (armchair). We also calculated the strain-stress relation for three-, four-layer and bulk black phosphorus and found their strain-stress behavior is similar to that of two-layer structure. They all have the critical strains of 24% and 32% in the x and y directions, respectively, and similar ideal strengths as the two-layer system.

To understand this enormous critical strain, we examined its crystal structure under 30% tensile strain and compared it to the relaxed configuration. The parameters, such as bond lengths, bond angles, and dihedral angles are presented in Table 1. The P-P bond lengths $r_1$ and $r_2$, bond angles $\alpha$ and $\beta$, the distance d between the puckered layers of the phosphorene in the z direction are denoted in Figure 1. $\phi_{1234}$ and $\phi_{1235}$ are the dihedral angles of the atoms 1-2-3-4, and 1-2-3-5, respectively. For the relaxed phosphorene, the bond lengths are $r_1$ = 2. 22 Å and $r_2$ = 2.26 Å. The puckered-layer distance d = 2.51 Å. The bond angles $\alpha$ and $\beta$ are 95.9° and 104.1°, respectively. The dihedral angles $\phi_{1234}$ = 73.8° and $\phi_{1235}$ = 36.7°. Under the axial tensile strain loaded in the zigzag direction, it was found that $r_1$, d and $\alpha$ experience significant changes. For example, the angle $\alpha$ between atoms 1, 2 and 3 in the xy plane opens from 95.9° to 117.5°, corresponding to 22.5% increase. The bond length $r_1$ lying in the xy plane is stretched from 2.22 Å to 2.50 Å, which is 12.6% extension compared to that of relaxed phosphorene. The pucker-layer distance d is reduced from 2.51 Å to 2.13 Å, (15.1% reduction). However, the bond length $r_2$ and angle $\beta$, which are not lying in the xy plane demonstrate smaller changes, with 3.5% and 7.4% reduction, respectively. In addition, the change percentages of the dihedral angles are both less than 8%.

**Table 1. The bond lengths, puckered-layer distance, bond angles and dihedral angles of the relaxed and strained monolayer phosphorene. The bond lengths $r_1$, $r_2$, puckered-layer distance d, bond angles $\alpha$ and $\beta$ are described in Figure 1. $\phi_{1234}$ and $\phi_{1235}$ are the dihedral angles of the atoms 1-2-3-4 and 1-2-3-5, respectively. The values in the parentheses are the change percentages compared to that of relaxed phosphorene.**

| system | $r_1$ (Å) | $r_2$ (Å) | d (Å) | $\alpha$ (°) | $\beta$ (°) | $\phi_{1234}$ (°) | $\phi_{1235}$ (°) |
|---|---|---|---|---|---|---|---|
| relaxed | 2.22 | 2.26 | 2.51 | 95.9 | 104.1 | 73.8 | 36.7 |
| $\varepsilon_x = 30\%$ | 2.50 (12.6%) | 2.18 (-3.5%) | 2.13 (-15.1%) | 117.5 (22.5%) | 96.4 (-7.4%) | 79.3 (7.5%) | 39.1 (6.5%) |
| $\varepsilon_y = 30\%$ | 2.21 (-0.5%) | 2.40 (6.2%) | 1.89 (-24.7%) | 94.0 (-2.0%) | 115.0 (10.5%) | 60.0 (-18.7%) | 29.8 (-18.8%) |

However, a different situation occurs for the 30% strain applied in the armchair direction. The bond lengths and the angle α show relatively small changes (no larger than 6.2%). In contrast, the puckered-layer distance d and dihedral angles $\phi_{1234}$ and $\phi_{1235}$ experience significant reduction. The distance d is reduced from 2.51 Å to 1.89 Å, which is 24.7 % reduction. Both dihedral angles are reduced by about 19%. This implies that the tensile strain in the armchair direction *effectively flattens* the pucker of phosphorene, rather than extensively extending the P-P bond lengths and opening the bond angles.

Meanwhile, the structural changes presented in Table 1 can also be used to explain the different ideal strengths (and critical strains) along the zigzag and armchair directions shown in Figure 2(a). The ideal strength in the zigzag direction is more than doubled compared to that in the armchair direction (18 GPa versus 8 GPa). Moreover, the Young's modulus in the zigzag direction is over three times larger than along the armchair axis. The lower stiffness in the armchair direction results from the smaller alteration of the in-plane crystal structure under the same amount of tensile strain. For example, a 30% tensile strain in the armchair direction is primarily to stretch the pucker of phosphorene, shown by the large changes in d, β, $\phi_{1234}$ and $\phi_{1235}$ listed in Table 1. The modification of the xy in-plane structural parameters, such as the bond length $r_1$ and angle α, are negligible (0.5% and 2.0% reduction, respectively). These compromised dihedral angles dramatically reduce the required strain energy.

In addition to the ideal tensile strength and critical stain, we also calculated other mechanical properties of the material, such as elastic constants, Young's and shear moduli, and Poisson's ratios. Since our systems are 2D structures, the elastic constants and moduli are from the Hooke's law under plane-stress condition [44],

$$\begin{bmatrix} \sigma_{xx} \\ \sigma_{yy} \\ \sigma_{xy} \end{bmatrix} = \frac{1}{1-v_{xy}v_{yx}} \begin{bmatrix} E_x & v_{yx}E_x & 0 \\ v_{xy}E_y & E_y & 0 \\ 0 & 0 & G_{xy}(1-v_{xy}v_{yx}) \end{bmatrix} \begin{bmatrix} \varepsilon_{xx} \\ \varepsilon_{yy} \\ 2\varepsilon_{xy} \end{bmatrix} = \begin{bmatrix} C_{11} & C_{12} & 0 \\ C_{21} & C_{22} & 0 \\ 0 & 0 & C_{66} \end{bmatrix} \begin{bmatrix} \varepsilon_{xx} \\ \varepsilon_{yy} \\ 2\varepsilon_{xy} \end{bmatrix} \quad (1)$$

Where $E_i = \frac{\sigma_i}{\varepsilon_i}$ is the Young's modulus along the axis of $i$, $v_{ij} = -\frac{d\varepsilon_j}{d\varepsilon_i}$ is the Poisson's ratio with tensile strain applied in the direction $i$ and the response strain in the direction $j$, $G_{xy}$ is the shear modulus on the xy plane. Based on Equation (1), the relation between the Young's and shear moduli, Poisson's ratios and elastic stiffness constants for a 2D system can be derived,

$$E_x = \frac{C_{11}C_{22} - C_{12}C_{21}}{C_{22}}, \quad E_y = \frac{C_{11}C_{22} - C_{12}C_{21}}{C_{11}}, \quad \nu_{xy} = \frac{C_{21}}{C_{22}}, \quad \nu_{yx} = \frac{C_{12}}{C_{11}}, \quad G_{xy} = C_{66}. \tag{2}$$

In order to calculate the elastic stiffness constants, the moduli, and Poisson's ratios in Equation (2), we scanned the energy surface of the materials in the strain range $-2\% < \varepsilon_{xx} < +2\%$ and $-2\% < \varepsilon_{yy} < +2\%$ with an increment of 0.5%, and $-1\% < \varepsilon_{xy} < +1\%$. The strain energy was calculated as $E_s = E(\varepsilon) - E_0$, where $E(\varepsilon)$ and $E_0$ are the total energy of strained and relaxed system, respectively. Since the strain energy is quadratic dependent on the applied strains,

$$E_s = a_1 \varepsilon_{xx}^2 + a_2 \varepsilon_{yy}^2 + a_3 \varepsilon_{xx} \varepsilon_{yy} + a_4 \varepsilon_{xy}^2 \tag{3}$$

Through parabolic fitting of the strain-energy surface, the coefficients $a_i$ in Equation (3) can be determined. In addition, the elastic stiffness constants can be calculated using the formula,

$$C_{ij} = \frac{1}{A_0 d_0} \left( \frac{\partial E_s^2}{\partial \varepsilon_i \partial \varepsilon_j} \right), \tag{4}$$

where $i, j = xx, yy,$ or $xy$, $A_0$ is the area of the simulation cell in the xy plane and $d_0$ is the effective thickness of the system.

From Eqs (3) and (4), the relations between the elastic stiffness constants $C_{ij}$ and the coefficients $a_i$ can be found. Consequently, we can derive the Young's/shear moduli and Poisson's ratios for our 2D systems as a function of $a_i$ as follows,

$$E_x = \frac{4a_1 a_2 - a_3^2}{4a_2 A_0 d_0}, \quad E_y = \frac{4a_1 a_2 - a_3^2}{4a_1 A_0 d_0}, \quad G_{xy} = \frac{2a_4}{A_0 d_0}, \quad \nu_{xy} = \frac{a_3}{2a_2}, \quad \nu_{yx} = \frac{a_3}{2a_1} \tag{5}$$

Our calculated moduli and Poisson's ratios are listed in Table 2. Due to the anisotropicity of the phosphorene structure, the Young's modulus (Poisson's ratio) has different value in the zigzag and armchair directions. For the monolayer phosphorene, the Young's modulus and Poisson's ratio in the zigzag direction are 3.8 times larger than their counterpart in the armchair direction. For the multi-layer phosphorene, their values in the zigzag direction are 4.3 times larger than their counterparts in the armchair direction, indicating it is more difficult to apply strain in the zigzag than the armchair direction.

**Table 2. The calculated moduli and Poisson's ratios for a monolayer phosphorene and few-layer black phosphorus.**

| System | Young's modulus (GPa) | | Shear modulus (GPa) | Poisson's ratio | |
|---|---|---|---|---|---|
| | [100] | [010] | | [100] | [010] |
| **1-layer** | 166 | 44 | 41 | 0.62 | 0.17 |
| **2-layer** | 162 | 38 | 43 | 0.71 | 0.17 |
| **3-layer** | 160 | 37 | 44 | 0.73 | 0.17 |
| **4-layer** | 159 | 37 | 45 | 0.73 | 0.17 |

Compared to other 2D materials such as graphene, $MoS_2$, BN etc, phosphorene demonstrates super flexibility with a much smaller Young's modulus. For example, the reported Young's moduli for graphene, $MoS_2$, and BN, are 1.0 TPa [45], 0.33 TPa [46], 0.25 TPa [47], respectively, compared to 0.166 TPa (zigzag) and 0.044 TPa (armchair) for phosphorene. This smaller Young's modulus in phosphorene may be resulted from two aspects: (1) weaker P-P bond strength and (2) the compromised dihedral angles rather than bond length stretch when a tensile strain is applied. This makes phosphorene in a great choice for practical large-magnitude-strain engineering.

Unlike the isotropic Young's modulus for graphene and $MoS_2$, the Young's modulus of phosphorene is sensitively dependent on the direction. To further explore the Young's modulus along an arbitrary direction, we derived the following equation for a 2D system,

$$\frac{1}{E_\varphi} = S_{11}\cos^4\varphi + (2S_{12} + S_{66})\cos^2\varphi\sin^2\varphi + S_{22}\sin^4\varphi \tag{6}$$

Where $\varphi \in [0, 2\pi]$ is the angle of an arbitrary direction from the +x axis, $E_\varphi$ is the Young's modulus along that direction, $S_{ij}$ are elastic compliance constants, which are correlated to elastic stiffness constants,

$$S_{11} = \frac{C_{22}}{C_{11}C_{22} - C_{12}^2}, S_{22} = \frac{C_{11}}{C_{11}C_{22} - C_{12}^2}, S_{12} = -\frac{C_{12}}{C_{11}C_{22} - C_{12}^2}, S_{66} = \frac{1}{C_{66}} \tag{7}$$

Note that $C_{12} = C_{21}$ and $S_{12} = S_{21}$. The direction dependence of the Young's modulus in the monolayer phosphorene is presented in Figure 3. The maximal Young's modulus of 166 GPa is along the zigzag direction and the minimal value of 44 GPa is along the armchair direction. The modulus in other arbitrary directions has a value between 44 and 166 GPa. The average value of the Young's modulus among all directions is 94 GPa.

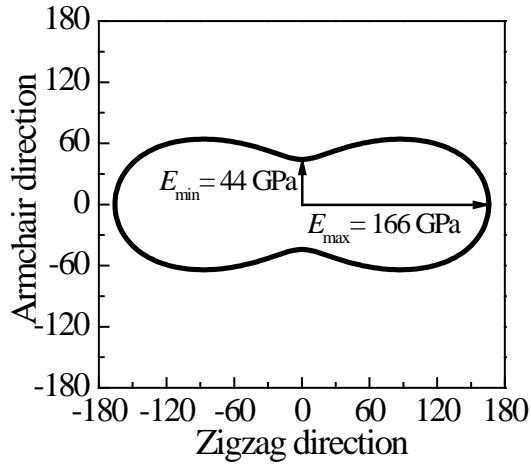

**Figure 2. The direction dependence of Young's modulus of a monolayer phosphorene.**

In summary, we investigated the mechanical properties of 2D monolayer phosphorene and few-layer black phosphorus through first principles DFT calculations. We found the materials demonstrated superior mechanical flexibility. Monolayer phosphorene can withstand stress up to 18 GPa and 8 GPa in the zigzag and armchair directions, respectively. It can hold critical strain up to 30%, while few-layer black phosphorus can sustain strain up to 32%. This strain limit of phosphorene is resulted from its unique puckered crystal structure. We found that the tensile strain applied in the armchair direction *effectively flattens* the pucker of phosphorene, rather than significantly extending the P-P bond lengths. These compromised dihedral angles significantly reduce the required strain energy. The much smaller Young's modulus in phosphorene, compared to other 2D materials, suggests great applications of large-magnitude-strain engineering. Moreover, the materials show strong anisotropicity and the calculated Young's modulus and Poisson's ratio in the zigzag directions are about four times larger than those in the armchair direction. We also derived a general model to calculate the Young's modulus in an arbitrary direction for 2D materials.

**Acknowledgement**

This work is supported by the Faculty Research Initiative Fund from School of Letters and Sciences at Arizona State University (ASU) to Peng. The authors thank ASU Advanced

Computing Center for providing resources. Dr. Fu Tang and Andrew Copple are acknowledged for the helpful discussions and critical review of the manuscript.

* To whom correspondence should be addressed.  E-mail: [xihong.peng@asu.edu](mailto:xihong.peng@asu.edu).

**Table caption**

**Table 1. The bond lengths, puckered-layer distance, bond angles and dihedral angles of the relaxed and strained monolayer phosphorene. The bond lengths $r_1$, $r_2$, puckered-layer distance d, bond angles $\alpha$ and $\beta$ are described in Figure 1. $\phi_{1234}$ and $\phi_{1235}$ are the dihedral angles of the atoms 1-2-3-4 and 1-2-3-5, respectively. The values in the parentheses are the change percentages compared to that of relaxed phosphorene.**

**Table 2. The calculated moduli and Poisson's ratios for a monolayer phosphorene and few-layer black phosphorus.**

**Figure captions**

**Figure 1,** Snapshots of a monolayer phosphorene structure. In (b), the dashed rectangle indicates a unit cell. Atoms were labeled using the numbers 1 through 6. Bond lengths $r_1$ and $r_2$, bond angles $\alpha$ and $\beta$ are denoted. The distance between the puckered layers in the z direction is labeled as d. (c) and (d) present the geometry of phosphorene under 30% axial tensile strain applied in the zigzag and armchair directions, respectively.

**Figure 2,** The strain-stress relation for (a) monolayer and (b) two-layer phosphorene structures. The monolayer can sustain stress up to 18 GPa and 8 GPa in the zigzag and armchair directions, respectively. The corresponding critical strains are 27% (zigzag) and 30% (armchair). The ideal strengths for the multi-layer phosphorene are 16 GPa and 7.5 GPa in the zigzag and armchair directions, respectively, and their critical strains are 24% (zigzag) and 32% (armchair).

**Figure 3,** The direction dependence of Young's modulus of a monolayer phosphorene.